\documentclass[prd,showpacs,amsmath,amssymb]{revtex4}
\usepackage{amsmath}
\usepackage{amsfonts}
\usepackage{mathrsfs}
\usepackage{pstricks}
\usepackage{color}
\usepackage{slashed}
\usepackage{dcolumn}% Align table columns on decimal point
\usepackage{epsfig,graphicx}
\newcommand{\be}{\begin{equation}}
\newcommand{\ee}{\end{equation}}
\newcommand{\ba}{\begin{array}{c}}
\newcommand{\ea}{\end{array}}
\newcommand{\bqa}{\begin{eqnarray}}
\newcommand{\eqa}{\end{eqnarray}}
\newcommand{\bm}[1]{\mbox{\boldmath{$#1$}}}
\makeatletter
    
    \newcommand{\Rmnum}[1]{\expandafter\@slowromancap\romannumeral #1@}
    \makeatother

\begin{document}

\bibliographystyle{unsrt}

\title{\bf  Study of the anomalous cross-section lineshape of $e^+e^-\to D\bar D$ at $\psi(3770)$
with an effective field theory}

\author{Guo-Ying Chen$^1$ and Qiang Zhao$^{2,3}$ }

\affiliation{1) Department of Physics, Xinjiang University, Urumqi
830046, China }

\affiliation{2) Institute of High Energy Physics, Chinese Academy of
Sciences, Beijing 100049, China }

\affiliation{3) Theoretical Physics Center for Science Facilities,
CAS, Beijing 100049, China}

\date{\today}

\begin{abstract}
We study the anomalous cross-section lineshape of $e^+e^-\rightarrow
D\bar D$ with an effective field theory. Near the threshold, most of
the $D\bar D$ pairs are from the decay of $\psi(3770)$. Taking into
account the fact that the nonresonance background is dominated by
the $\psi(2S)$ transition, the produced $D\bar D$ pair can undergo
final-state interactions before the pair is detected. We propose an
effective field theory for the low-energy $D\bar D$ interactions to
describe these final-state interactions and find that the anomalous
lineshape of the $D\bar D$ cross section observed by the BESII
collaboration can be well described.
\end{abstract}


\maketitle

%\par\vskip20pt
%\par

As the first charmonium state above the $D\bar D$ threshold, the
resonance $\psi(3770)$ is different from other charmonia with lower
masses. Because the $\psi(3770)$ decay into the open charm
$D\bar{D}$ is allowed by the Okubo-Zweig-Iizuka (OZI) rule, this
dominant decay mode leads to a broad width up to $27.2\pm 1.0$
MeV~\cite{PDG2012}. Obviously, the direct production process of
$e^+e^-\to \psi(3770)\rightarrow D\bar D$ is useful for the study of
the properties of $\psi(3770)$. In Ref.~\cite{BESPLB}, BESII
collaboration reported an anomalous behavior of the cross-section
lineshape at the $\psi(3770)$ mass region in $e^+e^-\rightarrow
D\bar D$ that cannot be described by a simple Breit-Winger of
$\psi(3770)$. Such an observation has inspired interesting
theoretical discussions~\cite{Zhao,Li,Liu,Achasov}. In particular,
it was found that the interfering effect between $\psi(3770)$ and
$\psi(2S)$ plays a very important role in understanding the
anomalous lineshape of $D\bar D$ at the $\psi(3770)$
resonance~\cite{Zhao,Li}. Such an interference can be recognized by
a relative phase factor $e^{i\phi}$, which is introduced between
these two resonances, and the phase angle $\phi$ must be large to
describe the anomalous $D\bar D$ lineshape.

In principle, the phase factor $e^{i\phi}$ can come from the
final-state interactions of $D\bar D$. Thus, it should be
interesting to study the $D\bar D$ anomalous lineshape using an
effective field theory to describe the $D\bar D$ final-state
interactions. This forms our motivation for this work. Near the
threshold, the $D\bar D$ pair produced in $e^+e^-\rightarrow D\bar
D$ comes from the decay of the $\psi(3770)$ and other nonresonance
background processes. Once the $D\bar D$ pair is produced, it could
undergo final-state interactions before it converts into the final
observed $D\bar D$ state. This phenomenon could explain the relative
phase between the $\psi(3770)$ and other non-$\psi(3770)$ amplitude
and provide a description of the $D\bar D$ lineshape. We note that
there are several cases in which the final-state interactions play
important roles in the understanding of the cross-section
lineshapes~\cite{CDM,CM,Mei1,Mei2}.

It is well known that an effective field theory is a useful tool to
study the low-energy hadron interactions. An effective field theory
utilizes the Tailor expansion of the small ratio between the typical
small scale $p$ and the cutoff scale $\Lambda$. For example, in
Chiral Perturbation Theory (ChPT), $p$ is the momentum of the
low-energy pion or pion mass, whereas $\Lambda=M_{\rho(770)}$ sets
the cutoff scale of this effective theory. An effective field theory
for the low-energy $D\bar D$ is different from that for the
low-energy $\pi \pi$ interaction because the $\psi(3770)$ should be
included explicitly into the effective Lagrangian. In addition to
the three-vector momentum of the $D$ ($\bar D$) meson, another small
scale, $\delta=M_{\psi(3770)}-2M_D\approx 40$ MeV, also appears in
the effective theory. This additional small scale will make the
power counting different from that in ChPT. A systematic development
of the effective field theory with resonances as intermediate states
is still under exploration, and interesting discussions on this
subject can be found in Refs.~\cite{Kaplan,Stefan}.

In this work, we use the effective field theory to study the $D\bar
D$ interaction to understand the dynamic details of the anomalous
cross-section lineshape observed by the BESII
Collaboration~\cite{BESPLB}.

At the beginning, we assume that the production of $D\bar D$ in
$e^+e^-$ annihilation can be approximated by the vector meson
dominance (VMD). This assumption means that the cross section for
$e^+e^-\rightarrow D\bar D$ is dominated by intermediate vector
meson productions via $e^+e^-\rightarrow \gamma^\ast \rightarrow
\mathcal{R}_i\rightarrow D\bar D$, where $\mathcal{R}_i$ denotes any
vector meson with an isospin of $I=0$ or $I=1$. However, it is
impossible to sum the contributions from all of the $\mathcal{R}_i$
in reality. As a reasonable approximation, one can include the
contributions from the vector mesons in the vicinity of the
considered energy region but neglect those far off-shell vector
mesons. In the energy region of the BES data from 3.74 GeV to 3.8
GeV, one can expect that $\psi(3770)$ plays the most important role
among all of the $\mathcal{R}_i$, whereas the contributions from all
the other $\mathcal{R}_i$ can be treated as background. As shown in
Ref.~\cite{Zhao}, the contribution from $\psi(2S)$ dominates the
background, whereas the contributions from other states are
negligible. Therefore, we only include the contributions from the
resonances $\psi(3770)$ and $\psi(2S)$ and neglect those from the
other resonances. Namely, $\psi(2S)$ would be the main background
near the $D\bar{D}$ threshold.

In VMD~\cite{Yennie,Ligang}, the coupling between the vector meson
and a virtual photon can be described as
\begin{equation}
\mathcal{L}_{V\gamma}=\frac{e M_V^2}{f_V}V_\mu A^\mu,
\end{equation}
where $V_\mu$ is the vector meson field, $A_\mu$ is the photon
field, and $M_V$ is the mass of the vector meson. Setting the
electron mass to $m_e\approx 0$, the coupling can be obtained as
\begin{equation}
\frac{e}{f_V}=\left[\frac{3\Gamma_{ee}}{\alpha
M_V}\right]^{1/2},\label{efv}
\end{equation}
where $\Gamma_{ee}$ is the electron-position decay width of $V_\mu$
and $\alpha=1/137$ is the fine-structure constant.

Once the $D\bar D$ pair is produced from the decay of the vector
meson $\psi(3770)$ or $\psi(2S)$, the pair can undergo final-state
interactions through the rescattering processes $D\bar D\rightarrow
D\bar D\rightarrow \cdots\rightarrow D\bar D$, which can be
described by the effective field theory. In the energy region of
interest, the three-vector momentum $p$ of the $D(\bar D)$ meson is
small. Thus, it is possible to construct an effective field theory
for the low-energy $D\bar D$ interactions by making use of the
expansion of the small momentum $p$. Because the mass of
$\psi(3770)$ is just above the threshold of $D\bar D$, we need to
include $\psi(3770)$ explicitly in the formulation. Near the
threshold, the $D$($\bar D$) meson can be treated as
nonrelativistic. Thus, the interaction Lagrangian for the $D\bar D$
system with the quantum number $J^{PC}=1^{--}$ can be constructed as
\begin{eqnarray}
\delta\mathcal{L}&=&\mathcal{L}_{\psi D\bar D}+\mathcal{L}_{(D\bar
D)^2}\nonumber\\
\mathcal{L}_{\psi D\bar D}&=&ig_{\psi D\bar D}\{D^\dagger\bm{\nabla}
\bar D-\bm{\nabla} D^\dagger\bar D\}\cdot\bm{\psi}+ig_{\psi D\bar
D}\{\bar D^\dagger\bm{\nabla}
D-\bm{\nabla} \bar D^\dagger D\}\cdot\bm{\psi},\nonumber\\
\mathcal{L}_{(D\bar D)^2}&=&f_1\{D^\dagger\bm{\nabla} \bar
D-\bm{\nabla} D^\dagger\bar D\}\cdot\{\bm{\nabla} \bar D^\dagger
D-\bar D^\dagger\bm{\nabla}
D\}\nonumber\\
&&+f_3\{D^\dagger\tau^i\bm{\nabla} \bar D-\bm{\nabla}
D^\dagger\tau^i\bar D\}\cdot\{\bm{\nabla} \bar D^\dagger\tau^i
D-\bar D^\dagger\tau^i\bm{\nabla}
D\}+\cdots,\nonumber\\
\mbox{with} \ \ \ D&=&\left(
      \begin{array}{c}
        D^0 \\
        D^+ \\
      \end{array}
    \right)
, \ \ \ \bar D=\left(
                           \begin{array}{c}
                             \bar{D}^0 \\
                             D^- \\
                           \end{array}
                         \right),\label{Lag}
\end{eqnarray}
where $\psi$ is the field operator of $\psi(3770)$; $D$ ($\bar
D^\dagger$) annihilates a $D$($\bar D$) meson; $D^\dagger$ ($\bar
D$) creates a $D$($\bar D$) meson; $\tau^i$ is the Pauli matrix, and
the ellipsis denotes other contact terms with more derivatives that
are higher order terms. The first term in $\mathcal{L}_{(D\bar
D)^2}$ accounts for the interaction in the isospin singlet channel,
and the second term accounts for the isospin triplet channel. The
contributions from other resonances, which are not included in the
Lagrangian, can be saturated into the contact terms
$\mathcal{L}_{(D\bar D)^2}$. Therefore, we take the coefficients
such as $f_1, f_3$ to be complex, where the imaginary parts of these
terms come from the width of the saturated resonances and the $D\bar
D$ annihilation effect. With isospin symmetry, we only have to
consider the terms for the isospin singlet channel in
$\mathcal{L}_{(D\bar D)^2}$ to study the $D\bar D$ final-state
interactions because the $D\bar D$ pair comes from the decay of
$\psi(3770)$ and $\psi(2S)$ in our approach.

\begin{figure}[hbt]
\begin{center}
  \includegraphics[width=8cm]{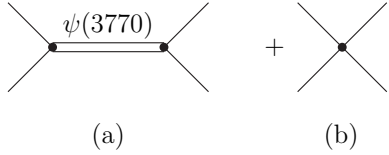}\\
  \caption{Tree diagrams for $D\bar D\rightarrow D\bar D$. (a) $D\bar D\rightarrow \psi(3770)\rightarrow D\bar D$, (b) contact interaction.}\label{tree}
  \end{center}
\end{figure}

\begin{figure}[hbt]
\begin{center}
  \includegraphics[width=12cm]{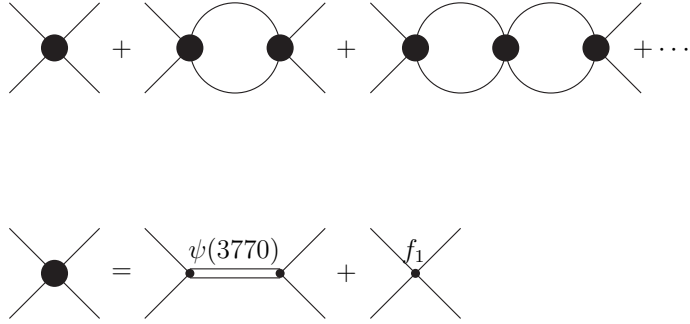}\\
  \caption{The bubble diagrams for the $D\bar D$ interactions where the potential is truncated at the leading order.}\label{bubble}
  \end{center}
\end{figure}

Now we come to the discussion of the power counting of this
effective field theory. The tree-level diagrams for the $D\bar D$
elastic scattering are shown in Fig.~\ref{tree}. Near the $D\bar D$
threshold, the denominator of the $\psi(3770)$ propagator can be
expressed as
\begin{eqnarray}
P(\psi)&=&\frac{1}{s-M_\psi^2+iM_\psi\Gamma_\psi^{\mbox{non-}D\bar D}}\nonumber\\
&\approx
&\frac{1}{(2M_D+p^2/M_D)^2-M_\psi^2+iM_\psi\Gamma_\psi^{\mbox{non-}D\bar D}}\nonumber\\
&=&\frac{1}{4p^2+4M_D^2-M_\psi^2+iM_\psi\Gamma_\psi^{\mbox{non-}D\bar
D}+\mathcal{O}(p^4)},
\end{eqnarray}
where $p$ is the magnitude of the three-vector momentum of the
$D$($\bar D$) meson in the overall center-of-mass frame,
$\Gamma_{\psi}^{\mbox{non-}D\bar D}$ denotes the non-$D\bar D$ decay
width of $\psi(3770)$, and $M_\psi$ is the mass of $\psi(3770)$. The
$D\bar D$ decay width of $\psi(3770)$ will be included through the
summation of the D meson loops in the following. Because
$\psi(3770)$ is close to the threshold of $D\bar D$, we expect that
$P(\psi)$ is at $\mathcal{O}(p^{-2})$. Taking the momentum power of
the $\psi D\bar D$ vertex into account, we find that
Fig.~\ref{tree}(a) is at $\mathcal{O}(p^0)$. From the naive power
counting, the leading contact terms have two derivatives; hence,
these terms are at $\mathcal{O}(p^2)$. However, in this naive power
counting, we have assumed that the coefficients of the contact
terms, i.e., $f_1,\ f_3,\cdots$, are at order of $\mathcal{O}(p^0)$.
In some cases, especially when there are bound states or resonances
near the threshold, the coefficients of the contact terms can be
enhanced. For example, in a $NN$ interaction, the S-wave contact
terms $C_S$ scale as $\mathcal{O}(p^{-1})$\cite{KSW}. Another
example is the $NN$ interaction in $^3P_0$, where the leading
contact term $C_{^3P_0}$ can be promoted to
$\mathcal{O}(p^{-2})$\cite{Long}. It is interesting to study whether
the same enhancement mechanism takes place in the $D\bar D$
interactions because the resonance $\psi(3770)$ is located near the
$D\bar{D}$ threshold. If $f_1$ is promoted to $\mathcal{O}(p^{-2})$
as $C_{^3P_0}$ in a $NN$ interaction, then the corresponding tree
diagram shown in Fig.~\ref{tree}(b) is at $\mathcal{O}(p^0)$, which
is the same as Fig.~\ref{tree}(a). However, because we do not know
the power of $f_1$ at the beginning, we then assume that the leading
contributions to $D\bar D$ elastic scattering come from both
Fig.~\ref{tree}(a) and (b). We will use the experiment data to
determine $f_1$ and see whether this contact term is enhanced.
Accordingly, the $D\bar D$ scattering amplitude in the specific
channel ($J^{PC}=1^{--}, I=0$) can be obtained by summing the bubble
diagrams as shown in Fig.~\ref{bubble}, which is equivalent to
solving the Lippmann-Schwinger equation $T=V+\int VGT$ with the
$D\bar D$ potential truncated at the leading order.

\begin{figure}[hbt]
\begin{center}
  \includegraphics[width=12cm]{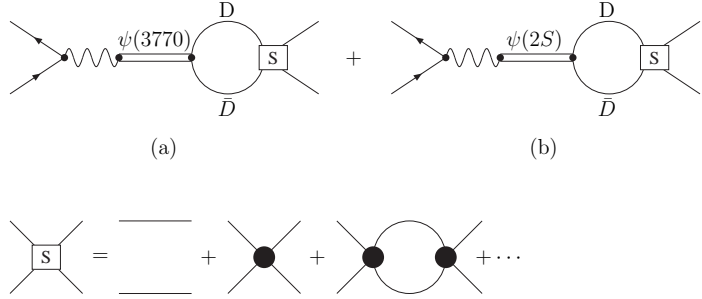}\\
  \caption{The Feynman diagrams for $e^+e^-\rightarrow D\bar D$ in our approach.}\label{feyndiagram}
  \end{center}
\end{figure}

Figure~\ref{feyndiagram} illustrates the final-state interactions
between the produced $D\bar D$. Because we first assume $f_1$ is
enhanced, which indicates the interaction between $D \bar D$ is
strong or the $D\bar D$ scattering length is large, we will use the
power divergent substraction (PDS) scheme proposed by
Ref.~\cite{KSW} to describe the large-scattering-length system in
our calculations. The loop integrals that we will encounter in
Fig.~\ref{feyndiagram} can generally be reduced to
\begin{eqnarray}
\mathcal{I}&\equiv&(\mu/2)^{4-D}\int\frac{d^D\ell}{(2\pi)^D}\frac{\vec{\ell}^2}{[\ell^0-\vec{\ell}^2/2M_D+i\epsilon]\cdot[E-\ell^0-\vec{\ell}^2/2M_D+i\epsilon]}\nonumber\\
&=&-i(\mu/2)^{4-D}\int\frac{d^{(D-1)}\ell}{(2\pi)^{(D-1)}}\frac{\vec{\ell}^2}{E-\vec{\ell}^2/M_D+i\epsilon}\nonumber\\
&=&iM_D(M_D E)(-M_D
E-i\epsilon)^{(D-3)/2}\Gamma(\frac{3-D}{2})\frac{(\mu/2)^{4-D}}{(4\pi)^{(D-1)/2}},
\end{eqnarray}
where $E=p^2/M_D$ is the total kinematic energy of the $D\bar D$
system. It is clear that this result is convergent in $D=4$ but
divergent in $D=3$. With the PDS scheme, we have to remove the $D=3$
pole in the above result by adding the counterterm
\begin{equation}
\delta\mathcal{I}=i\frac{M_D(M_D E)\mu}{4\pi(D-3)} .
\end{equation}
Hence the subtracted integral in $D=4$ reads
\begin{equation}
\mathcal{I}^{PDS}=\mathcal{I}+\delta\mathcal{I}=i\frac{M_D
}{4\pi}p^2(ip+\mu).
\end{equation}
Notice that $\mathcal{I}^{PDS}=\mathcal{I}$ at $\mu=0$, which is
simply the result in the minimal subtraction (MS) scheme. We can
choose $\mu$ to be the typical momentum scale of the $D(\bar D)$
meson, which is $p\leq 300$ MeV in our calculations.

We can then write down the amplitude for $e^+e^-\rightarrow D\bar D$
as
\begin{equation}
i\mathcal{M}=i\mathcal{M}_a+i\mathcal{M}_b.
\end{equation}

To be more specific, the amplitude for process
Fig.~\ref{feyndiagram}(a) reads
\begin{eqnarray}
i\mathcal{M}_a&=&-ie^2\bar v(k_2)\bm{\gamma} u(k_1)\cdot({\bm
p}_1-{\bm
p}_2)\frac{1}{s}\frac{M_\psi^2}{f_\psi}\frac{1}{s-M_{\psi}^2+iM_\psi
G_\psi}g_{\psi D\bar D},
\end{eqnarray}
with
\begin{eqnarray}
G_{\psi}&=&\Gamma_\psi^{\mbox{non-}D\bar D}+\frac{1}{12\pi
M_\psi}\left(g^2_{\psi D\bar
D}-f_1(s-M_\psi^2+iM_\psi\Gamma_\psi^{\mbox{non-}D\bar
D})\right)\left(\frac{|\vec{p}_{D^0}|^3-i|\vec{p}_{D^0}|^2\mu}{M_{D^0}}+\frac{|\vec{p}_{D^+}|^3-i|\vec{p}_{D^+}|^2\mu}{M_{D^+}}\right),\nonumber\\
\end{eqnarray}
where $\gamma$ is the Dirac gamma matrix; $k_1$ and $k_2$ are the
incoming momenta of the electron and positron, respectively, and
$p_1$ and $p_2$ are the outgoing momenta of $D$ and $\bar D$,
respectively. $G_\psi$ cannot be simply interpreted as the width of
$\psi(3770)$ because this term is a complex number. If we set
$f_1=0$ and $\mu=0$, then
$G_\psi=\Gamma_\psi=\Gamma_\psi^{\mbox{non-}D\bar
D}+({|\vec{p}_{D^0}|^3}/{M_{D^0}}+{|\vec{p}_{D^+}|^3}/{M_{D^+}}){g^2_{\psi
D\bar D}}/{(12\pi M_\psi)}$.

The amplitude for Fig.~\ref{feyndiagram}(b) can be written as
\begin{eqnarray}
i\mathcal{M}_b&=&-ie^2\bar v(k_2)\bm{\gamma} u(k_1)\cdot({\bm
p}_1-{\bm
p}_2)\frac{1}{s}\frac{M_{\psi(2S)}^2}{f_{\psi(2S)}}\frac{1}{s-M_{\psi(2S)}^2+iM_{\psi(2S)}
\Gamma_{\psi(2S)}}\tilde{g}_{\psi(2S)},
\end{eqnarray}
with
\begin{eqnarray}\tilde{g}_{\psi(2S)}&=&g_{\psi(2S)D\bar
D}\left[1+i\frac{1}{12\pi }\left(-f_1+\frac{g_{\psi D\bar
D}^2}{s-M_\psi^2+i M_\psi\Gamma_{\psi}^{\mbox{non-}D\bar
D}}\right)\left(\frac{|\vec{p}_{D^0}|^3-i|\vec{p}_{D^0}|^2\mu}{M_{D^0}}+\frac{|\vec{p}_{D^+}|^3-i|\vec{p}_{D^+}|^2\mu}{M_{D^+}}\right)\right]^{-1},\nonumber\\
\end{eqnarray}
where the PDG~\cite{PDG} value for the $\psi(2S)$ mass
$M_{\psi(2S)}$ can be adopted, and $f_{\psi(2S)}$ can be extracted
by Eq.~(\ref{efv}) using $\Gamma_{\psi(2S)\rightarrow e^+e^-}=2.35$
keV~\cite{PDG}.

To proceed, we denote the cross section for $e^+e^-\rightarrow D\bar
D$ as $\sigma^B(s)$, which does not include the initial state
radiation (ISR) effect. In reality, for a given energy $\sqrt{s}$,
the actual c.m. energy for the $e^+e^-$ annihilation is
$\sqrt{s^\prime}=\sqrt{s(1-x)}$ due to the ISR effect, where
$xE_{\mbox{beam}}$ is the total energy of the emitted photons. To
order $\alpha^2$ radiative correction in the $e^+e^-$ annihilation,
the observed cross section $\sigma^{\mbox{obs}}$ at BESII can be
related to our result $\sigma^B$ through~\cite{ISR}
\begin{eqnarray}\label{xsect-isr}
\sigma^{\mbox{obs}}(s)=(1+\delta_{VP})\int_0^{1-4M_D^2/s}dx
f(x,s)\sigma^B(s(1-x)),
\end{eqnarray}
where $(1+\delta_{VP})=1.047$, and the function $f(x,s)$ is given by
\begin{eqnarray}
f(x,s)&=&\beta x^{\beta-1}\delta^{V+S}+\delta^H,\nonumber\\
\beta&=&\frac{2\alpha}{\pi}(\ln\frac{s}{m_e^2}-1),\nonumber\\
\delta^{V+S}&=&1+\frac{3}{4}\beta+\frac{\alpha}{\pi}(\frac{\pi^2}{3}-\frac{1}{2})+\frac{\beta^2}{24}(\frac{1}{3}\ln\frac{s}{m_e^2}+2\pi^2-\frac{37}{4}),\nonumber\\
\delta^H&=&-\beta(1-\frac{x}{2})+\frac{1}{8}\beta^2\left[4(2-x)\ln\frac{1}{x}-\frac{1+3(1-x)^2}{x}\ln(1-x)-6+x\right].
\end{eqnarray}

Before fitting the BESII data with Eq.~(\ref{xsect-isr}), we first
discuss our treatment of $\Gamma_\psi^{\mbox{non-}D\bar D}$. It
seems impossible to determine $\Gamma_\psi^{\mbox{non-}D\bar D}$
definitely in our fitting because $\Gamma_\psi^{\mbox{non-}D\bar D}$
is always accompanied by $f_1$ in our formula, and any change of
$\Gamma_\psi^{\mbox{non-}D\bar D}$ can be compensated by tuning
$f_1$. The experimental results on the non-$D\bar D$ branching ratio
of $\psi(3770)$ decay are still
controversial~\cite{CLEO,BES,Anashin:2011kq}. In contrast, the
next-to-leading-order (NLO) pQCD calculation expects the
non-$D\bar{D}$ decay branching ratio to be at most approximately
$5\%$~\cite{HFC}. Meanwhile, an effective Lagrangian approach
estimates that the $D$ meson loop rescatterings into non-$D\bar{D}$
light vector and pseudoscalar mesons leads to approximately 1\%
non-$D\bar{D}$ branching ratios~\cite{Zhang:2009kr}. A similar
calculation by Ref.~\cite{Liu:2009dr} also confirms such a
nonperturbative phenomenon. One also notices that so far, most of
the well-measured non-$D\bar D$ decay modes of $\psi(3770)$ are
found to be rather small. Namely, their branching ratios are either
at the order of $10^{-3}-10^{-4}$, or only an upper limit is
set~\cite{PDG}.

Taking all these facts into account and for the purpose of studying
the dominant $D\bar{D}$ channel, we set
$\Gamma_\psi^{\mbox{non-}D\bar D}$ to be zero in our fitting as a
leading approximation. We have checked that the fitting results are
approximately unchanged even though we set the non-$D\bar D$
branching ratio of the $\psi(3770)$ decay to be at the order of
several percent.

\par

\begin{table}[h]
\begin{center}
\begin{tabular}{|c||c|c|c|c|}
\hline       & MS($\mu=0$)  & PDS($\mu=\delta$)  & PDS($\mu=m_\pi$)   & PDS($\mu=300$MeV)\\
\hline \hline
$M_\psi(\mbox{GeV})$    &  $ 3.7674\pm 0.0044           $  &  $3.7685\pm 0.0056           $   &  $3.7725\pm0.0046           $ &  $3.7755\pm0.0041$\\
$e/f_\psi$              &  $ (4.25\pm0.13)\times 10^{-3}$  &  $(4.25\pm0.69)\times 10^{-3}$   &  $(4.75\pm0.61)\times10^{-3}$ &  $(5.48\pm0.67)\times10^{-3}$\\
$g_{\psi D\bar D}$      &  $ 15.4\pm2.7                 $  &  $14.6\pm2.5                 $   &  $12.5\pm2.0                $ &  $10.1\pm1.4$\\
$g_{\psi(2S)D\bar D}$   &  $  -6.9\pm3.6                $  &  $-6.3\pm4.3                 $   &  $-6.6\pm4.1                $ &  $-6.8\pm3.9$\\
$f_1(\mbox{GeV}^{-2})$  &  $  (2059\pm534)+i(0\pm836)   $  &  $(2096\pm504)+i(76\pm935)   $   &  $(1871\pm553)+i(570\pm726) $ &  $(1288\pm452)+i(802\pm297)$\\
\hline
$\chi^2/d.o.f$    & $23.4/22\approx1.06$   & $23.3/22\approx1.06$& $23.3/22\approx 1.06$ & $23.4/22\approx 1.06$\\
\hline
\end{tabular}
\caption{Fitted parameters and fitting qualities with different
$\mu$. Here, we use $\delta=40$ MeV. }
\end{center}
\end{table}

The fitted parameters and fitting qualities with $\mu=\delta, \
m_\pi, \ 300$ MeV are shown in Table~\Rmnum{1}. For comparison, we
also show the result with $\mu=0$, which corresponds to the value in
the MS scheme. The result shows that the fitted parameters are
insensitive to the choice of $\mu$. Moreover, the real part of $f_1$
is large, at the order of $(M_D/\delta)^2$, which is consistent with
our previous assumption. In contrast, the imaginary part of $f_1$ is
not well determined. Note that the NLO term $f_1$ has a comparable
magnitude to that of the leading order term. This result suggests
that the effective field theory expansion may not be convergent.
Thus, the fitting results may not be quantitatively reliable. To
have a better understanding of our results, we investigate the
dependence of $f_1$ on the scattering length $a$ as that was done in
Ref.~\cite{KSW}.  For the $P$-wave $D\bar D$ elastic scattering, we
denote the Feynman amplitude as $i\mathcal{A}\cos\theta$, where
$\theta$ is the angle between the incoming and outgoing momenta in
the c.m. frame. Then, the correlation between $\mathcal{A}$ and the
$P$-wave phase shift $\delta$ is
\begin{equation}
\mathcal{A}=\frac{48\pi M_D p^2}{p^3\cot\delta-ip^3}.
\end{equation}
With the effective range expansion
\begin{equation}
p^{2\ell+1}\cot\delta_\ell(p)=\frac{-1}{a_\ell}+\frac{r_\ell}{2}p^2+\mathcal{O}(p^4),
\end{equation}
and taking the case of $P$-wave scattering ($\ell=1$), we then
obtain
\begin{equation}
\mathcal{A}=\frac{48\pi M_D
p^2}{-\frac{1}{a}+\frac{r_0}{2}p^2-ip^3+\mathcal{O}(p^4)}.\label{eq1}
\end{equation}
For simplicity and only illustrating some aspects of the effective
field theory, we ignore the $\psi(3770)$ and consider a $D\bar D$
effective theory with only the contact terms. Accordingly, the
tree-level amplitude for the $P$-wave scattering can be written as
\begin{equation}
i\mathcal{A}_{tree}=i\sum_{n=1}^\infty C_{2n}p^{2n}.
\end{equation}
For the isospin $I=0$ channel, we have the coefficient of the
leading contact term $C_2=8f_1$. The full amplitude can then be
obtained by summing over all the bubble diagrams as shown in
Fig.~\ref{bubble}. The amplitude becomes
\begin{equation}
\mathcal{A}=\frac{\sum C_{2n}p^{2n}}{1-\frac{ip+\mu}{48\pi M_D}\sum
C_{2n}p^{2n}}.\label{eq2}
\end{equation}
Using the fact that the amplitude $\mathcal{A}$ should be
independent of the arbitrary subtraction scale $\mu$, we can
determine the $\mu$ dependence of the coupling constants
$C_{2n}(\mu)$
\begin{equation}
\frac{dC_{2n}}{d\mu}=-\frac{1}{48\pi
M_D}\sum_{m=1}^{n-1}C_{2(n-m)}C_{2m}.
\end{equation}
Note that $\frac{dC_2}{d\mu}=0$. One can see that, different from
the $S$-wave scattering that was considered in Ref.~\cite{KSW}, the
coefficient of the leading contact term $C_2$ is independent of
$\mu$ for the $P$-wave scattering. This fact makes the PDS approach
fail to improve the convergence of the effective field expansion for
the $P$-wave scattering. By comparing Eqs.~(\ref{eq1}) and
(\ref{eq2}) with each other, we obtain
\begin{equation}
C_2=-48\pi M_D  a.
\end{equation}
For the $I=0$ channel, we have $f_1=C_2/8=-6\pi M_D a$, which
suggests that $f_1$ can be large if the $P$-wave scattering length
$a$ is sizeable. It is also interesting to notice that, if $a\sim
\frac{1}{\Lambda p^2}$, by choosing $p=m_\pi$ and the cutoff scale
$\Lambda=1$ GeV, we will have $f_1\sim 1900 \ \mbox{GeV}^{-2}$,
which is close to our fitted value.

\begin{figure}[hbt]
\begin{center}
\includegraphics[width=6cm]{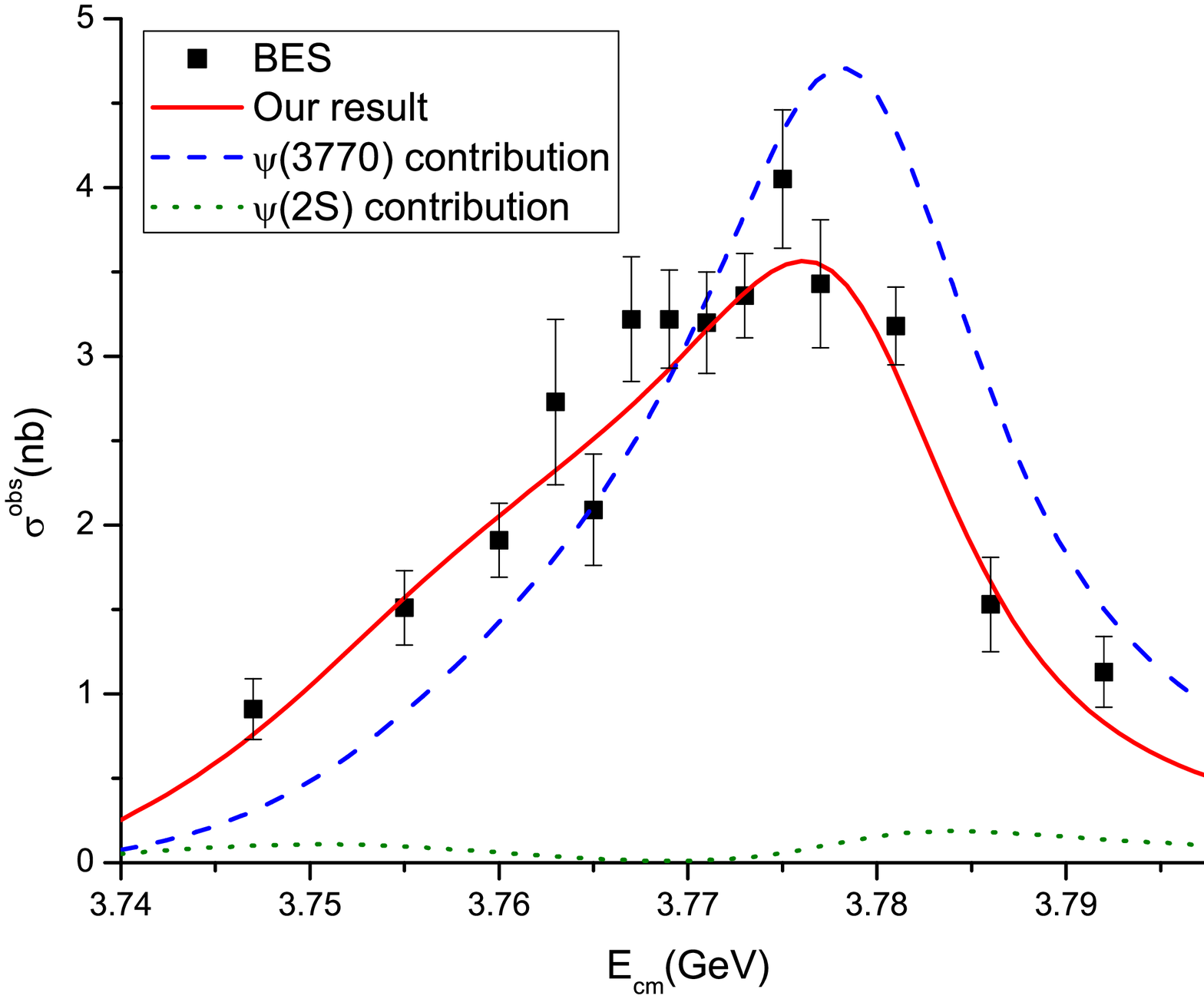}\ \ \ \ \ \ \ \ \includegraphics[width=6cm]{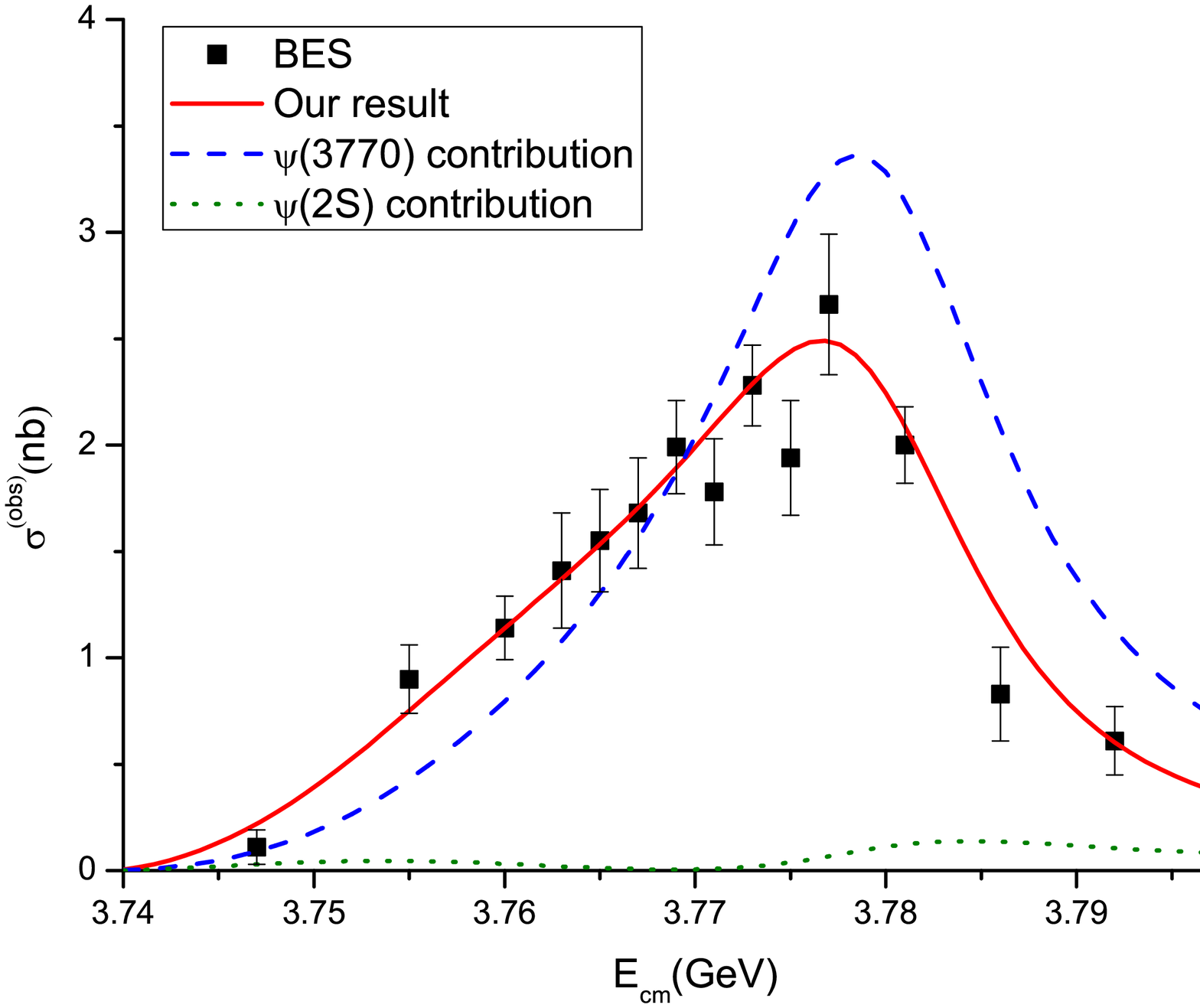}
\caption{The observed cross sections for $e^+e^-\rightarrow D^0\bar
D^0$ (left plot) and $e^+e^-\rightarrow D^+D^-$ (right plot) with
$\mu=\delta$. The solid line is the fitting result in our approach
shown in Fig.\ref{feyndiagram}, the dashed line shows the
contribution from Fig.\ref{feyndiagram}.(a), and the dotted line
shows the contribution from Fig.\ref{feyndiagram}.(b). The data are
from BES\cite{BESPLB}. }\label{fitdiagram}
  \end{center}
\end{figure}

Our fitting results for the $D\bar D$ cross-section lineshape are
presented in Fig.~\ref{fitdiagram}, where we only show the result
with $\mu=\delta$ because the other choice of $\mu$ gives similar
lineshapes.

With the fitted parameters, we can obtain the width and
electron-positron decay width of $\psi(3770)$ as the following:
\begin{eqnarray}
\Gamma_\psi&=&\frac{g_{\psi D\bar D}^2}{48\pi M_\psi^2}\left[(M^2_{\psi}-4M_{D^0}^2)^{3/2}+(M^2_{\psi}-4M_{D^+}^2)^{3/2}\right],\nonumber\\
\Gamma_{ee}&=&\frac{1}{3}\alpha M_\psi(\frac{e}{f_\psi})^2.
\end{eqnarray}
The corresponding values with different choices of $\mu$ are listed
in Table~\Rmnum{2}.

\begin{table}[h]
\begin{center}
\begin{tabular}{|c|c|c|c|c|}
\hline       & MS($\mu=0$)  & PDS($\mu=\delta$)  & PDS($\mu=m_\pi$)   & PDS($\mu=300$ MeV)\\
\hline
$\Gamma_\psi(\mbox{MeV})$    &  $ 27.5\pm 11.06           $  &  $25.9\pm10.9          $   &  $   22.4\pm8.2           $ &  $16.3\pm5.1               $\\
\hline
$\Gamma_{ee}(\mbox{eV}) $    &  $ 165.6\pm10.1          $  &  $165.6\pm53.8         $   &  $   207.1\pm53.2          $ &  $275.9\pm 67.5                   $\\
\hline
\end{tabular}
\caption{The total and electron pair decay widths determined by the
fitted parameters. }
\end{center}
\end{table}

\begin{figure}[h]
\begin{center}
  \includegraphics[width=6cm]{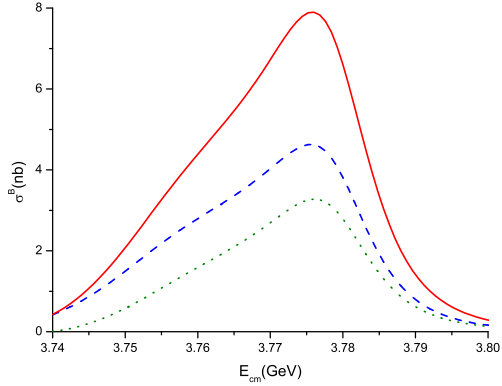}\\
  \caption{The Born cross section $\sigma^B(s)$ for $e^+e^-\rightarrow D\bar D$ with $\mu=\delta$. The solid line is for $e^+e^-\rightarrow
  D\bar D$, the dashed line is for $e^+e^-\rightarrow D^0\bar D^0$, and the dotted line is for $e^+e^-\rightarrow D^+D^-$.}\label{born}
  \end{center}
\end{figure}

In Fig.~\ref{born}, we also present the Born cross section
$\sigma^B(s)$ for $e^+e^-\rightarrow D\bar D$, which is denoted by
the solid curve. The dashed and dotted lines are for the neutral and
charged $D$-meson-pair Born cross sections, respectively. Combining
the results shown in Figs.~\ref{fitdiagram} and ~\ref{born}, we find
that the anomalous cross-section lineshape could originate from the
interferences from the $\psi(2S)$ pole and $D\bar{D}$ final-state
interactions. Because the $\psi(2S)$ pole is relatively isolated due
to its relatively narrow width in comparison with the mass gap
between $\psi(2S)$ and $\psi(3770)$, the relative phase between the
$\psi(2S)$ and $\psi(3770)$ amplitudes is likely to be produced by
the $D\bar{D}$ final-state interactions. Although our calculation
cannot determine the absolute value for the possible non-$D\bar{D}$
decay branching ratio of $\psi(3770)$, it is constructive to
recognize the important role played by the final-state $D\bar{D}$
interactions that cause the deviation of the $e^+e^-\to D\bar{D}$
cross section in the $\psi(3770)$ mass region from a Breit-Wigner
shape. This analysis is useful for our further understanding of the
$\psi(3770)$ non-$D\bar{D}$ decays as a manifestation of possible
nonperturbative QCD mechanisms.

To test the effect of the $f_1$ term, we can redo the fit by setting
$f_1=0$ in the MS scheme. The fitted parameters and fitting quality
are
\begin{eqnarray}
M_{\psi}&=&3.7844\pm 0.0012\ \ \mbox{GeV},\ \ \ \ \frac{e}{f_\psi}=(3.52\pm0.3)\times 10^{-3},\nonumber\\
g_{\psi D\bar D}&=&11.68\pm0.75,\ \ \ \ \ \ \ \ \ g_{\psi(2S) D\bar
D}=-(14.61\pm
1.35),\nonumber\\
\chi^2/\mbox{d.o.f.}&=&26.02/24\approx 1.08.
\end{eqnarray}
Note that $g_{\psi(2S)D\bar D}$ is more than two times larger than
our previous result.

\begin{figure}[hbt]
\begin{center}
\includegraphics[width=6cm]{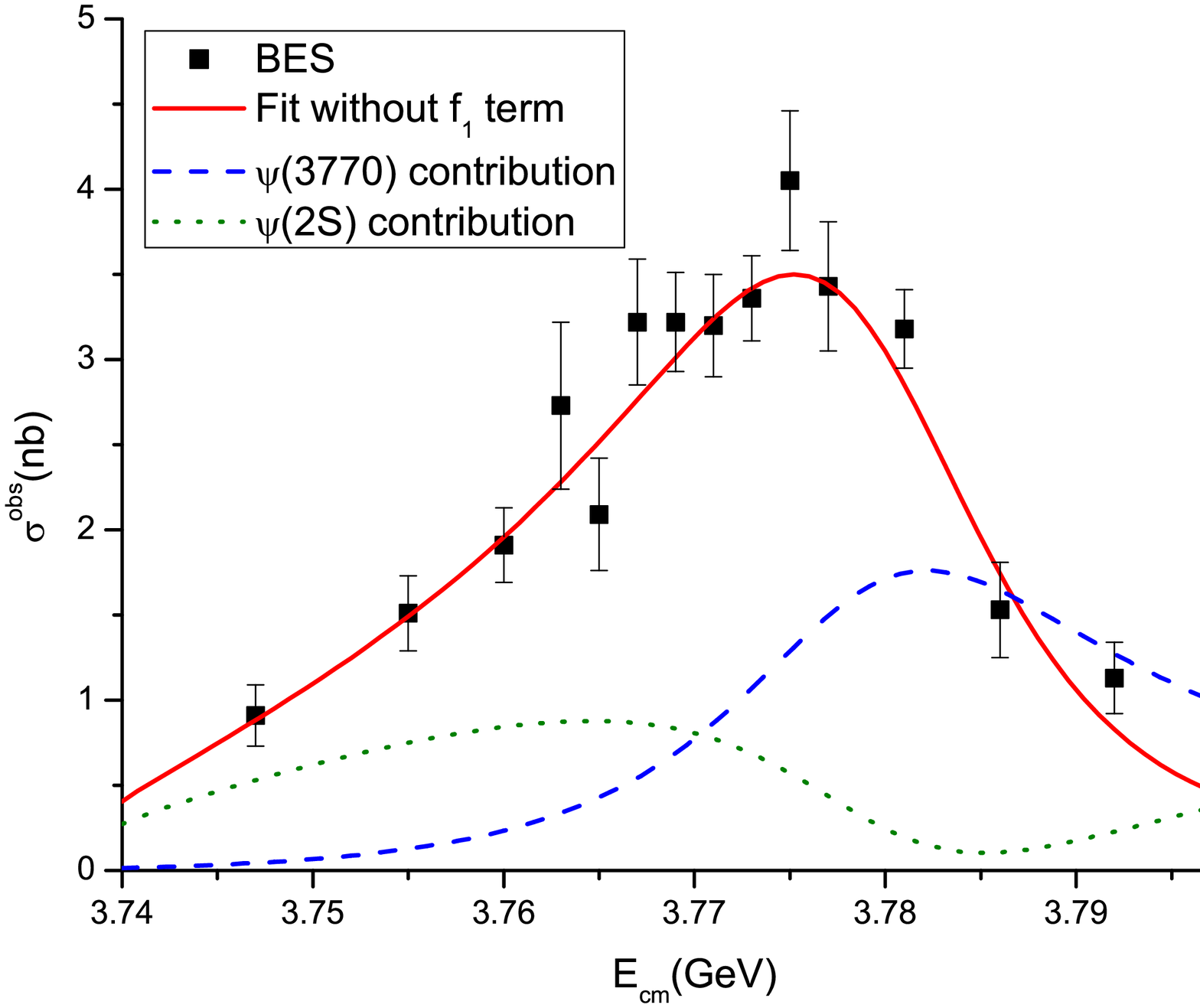}\ \ \ \ \ \ \ \ \includegraphics[width=6cm]{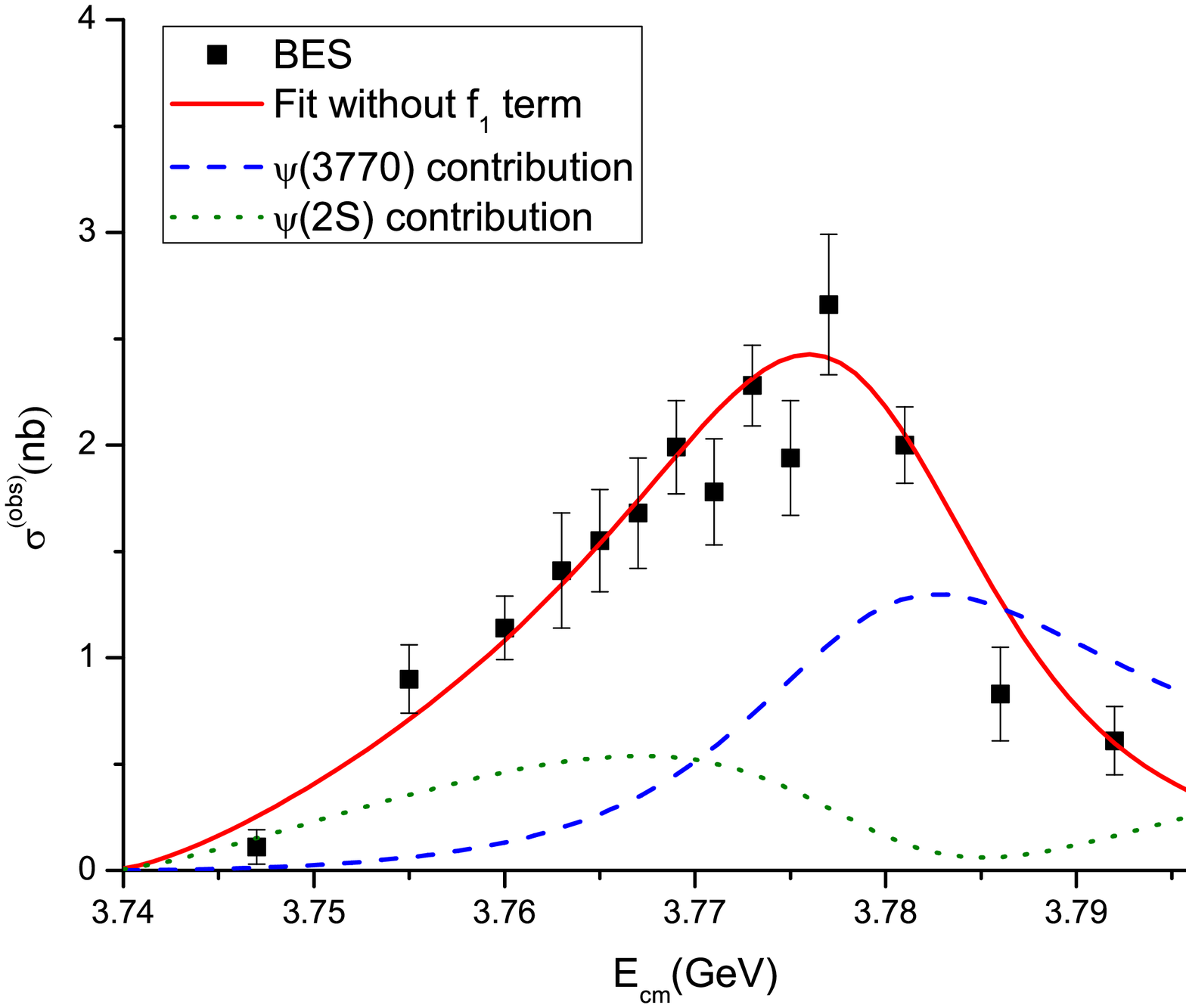}
\caption{The observed cross sections for $e^+e^-\rightarrow D^0\bar
D^0$ (left plot) and $e^+e^-\rightarrow D^+D^-$ (right plot). The
solid line is the fitting result with $f_1=0,$ and $\mu=0$, the
dashed line shows the contribution from $\psi(3770)$, and the dotted
line shows the contribution from $\psi(2S)$. The data are from
BES~\cite{BESPLB}. }\label{NOf1fitdiagram}
  \end{center}
\end{figure}

The fitted lineshape and exclusive contributions from $\psi(2S)$ and
$\psi(3770)$ are presented in Fig.~\ref{NOf1fitdiagram}.
Unsurprisingly, this figure shows that the contribution from
$\psi(2S)$ is larger than that in the previous fitting in
Fig.~\ref{fitdiagram} because of the larger coupling constant
$g_{\psi(2S) D\bar D}$. The distorted lineshape can be explained by
the interference between $\psi(3770)$ and $\psi(2S)$, which is
constructive at $E_{cm}<M_\psi$ but destructive at $E_{cm}>M_\psi$.
This observation can help us conclude that a large $g_{\psi(2S)D\bar
D}$ will favor a larger value for $M_\psi$, i.e., a larger mass for
$\psi(3770)$ than the present PDG average. We also find that, in
this fitting, the fitted $\chi^2$ is sensitive to $M_\psi$. For
example, the best fit gives $\chi^2\approx 41$ when we fix
$M_\psi=3.78$. By adopting the PDG values~\cite{PDG,PDG2012} for
$M_\psi$, the yield of $\chi^2$ can be even larger. Furthermore,
such a large $g_{\psi(2S)D\bar D}$ suggests that we need to include
the contact term $f_1$ in the $D\bar D$ interaction to saturate the
contribution from $\psi(2S)$. With this aspect taken into account,
we can affirm that the fitting result with $f_1=0$ is not
self-consistent. In general, the inclusion of the $f_1$ term seems
to be necessary to yield a reasonable value for $g_{\psi(2S)D\bar
D}$ and, at the same time, determine $M_{\psi}$ in a range closer to
the PDG average~\cite{PDG,PDG2012}.

In summary, we have proposed an effective field theory for
low-energy $D\bar D$ interactions in which we have included the
resonance $\psi(3770)$ and an additional small scale $\delta$. It is
found that the coefficient of the contact term $f_1$ will be
enhanced to be $\mathcal{O}(p^{-2})$. Therefore, the leading $D\bar
D$ interaction potential in this specific channel would come from
the $S$-channel $\psi(3770)$ exchange and the contact term $f_1$.
With the leading $D\bar{D}$ potential, we then sum the bubble
diagrams to describe the $D\bar D$ final-state interaction as shown
in Fig.~\ref{feyndiagram}. We find that we can describe the
anomalous cross-section lineshape of $e^+e^-\to D\bar D$ observed by
the BESII Collaboration~\cite{BESPLB} using the effective field
theory. This approach should be useful for our further understanding
of the $\psi(3770)$ non-$D\bar{D}$ decays, which could share the
same dynamic origin as the $D\bar{D}$ cross-section lineshape
anomaly as emphasized in Refs.~\cite{Zhang:2009kr,Zhao:2010ja}.

We also test the effects of the contact term $f_1$ and find that,
without this term, the extracted value of $g_{\psi(2S)D\bar D}$ is
too large to make the fitting self-consistent. Nevertheless, the
fitted $\psi(3770)$ mass is significantly larger than that in
PDG~\cite{PDG,PDG2012}. Our study also suggests that the
subthreshold $\psi(2S)$ plays an important role in our understanding
of the $D\bar D$ interactions. A better determination of
$g_{\psi(2S)D\bar D}$ should be strongly encouraged.

\section*{Acknowledgement}

We would like to thank Prof. J.-P. Ma, Dr. C. Meng and X.-S. Qin for
useful discussions.  This work is supported, in part, by National
Natural Science Foundation of China (Grant Nos. 11147022 and
11035006), Chinese Academy of Sciences (KJCX2-EW-N01), Ministry of
Science and Technology of China (2009CB825200), DFG and NSFC (CRC
110), and Doctor Foundation of Xinjiang University (No. BS110104).

\end{document}